\documentstyle[sprocl,epsf]{article}

\def\Journal#1#2#3#4{{#1} {\bf #2}, #3 (#4)}

\def\PLB{{\em Phys. Lett.}  B}

\def\be{\begin{equation}}
\def\ee{\end{equation}}
\def\bea{\begin{eqnarray}}
\def\eea{\end{eqnarray}}

\newcommand{\nc}{\newcommand}
\def\DBR#1#2{\bigl\{#1 \bigr\}\bigl\{#2 \bigr\}} 

\begin{document}


\title{THE QUANTUM BOLTZMANN EQUATION IN A NONTRIVIAL BACKGROUND}

\author{M. JOYCE}

\address{ LPT, Universit\'e Paris-XI, B\^atiment 211,
F-91405 Orsay Cedex, France
\\E-mail: Michael.Joyce@th.u-psud.fr}

\author{K. KAINULAINEN}

\address{NORDITA,
               Blegdamsvej 17, DK-2100, Copenhagen \O , Denmark
    \\E-mail: kainulai@nordita.dk}

\author{T. PROKOPEC}

\address{ Universit\'e de Lausanne, Institut de physique th\'eorique,
                     BSP, CH-1015 Lausanne, Suisse
  \\E-mail: Tomislav.Prokopec@ipt.unil.ch}

\maketitle

\abstracts{This talk\footnote{Talk delivered by T. Prokopec
at COSMO-99, Trieste, Italy, Sep 27 - Oct 2, 1999;
Report No. UNIL-IPT/00-02 (Feb 2000)}
is a status report on our study of
quantum transport equations relevant for baryogenesis computations.
Our main finding is that, as a consequence of localization in space,
the quasiparticle picture of the plasma dynamics breaks down at first
non-trivial order in gradient expansion. While in this talk we
focus on bosons, we expect that a similar picture holds for
fermions. We then argue that the quasiparticle picture is recovered in
the adiabatic limit of frequent scattering.
}

%
%
\section{Introduction and motivation}
\label{sec:Introduction and motivation}

A reliable computation of baryon production at the electroweak phase
transition requires a new formulation of the dynamics of the CP violating
sources in an out-of-equilibrium relativistic plasma. To this purpose we are
working on controlled derivation of the relativistic quantum transport
equations for weakly coupled plasma \cite{jkpP,jkp1,jkp2}.
These equations are essential to a systematic treatment of the
matter-antimatter creation at the electroweak phase transition, since they
describe the relevant CP violating sources, their transport and dissipation
in one formalism. Our aim is to apply this formalism to computation of the
baryon production at the electroweak transition in the Standard Model
and its supersymmetric extensions, and thus make contact with
the upcoming particle physics experiments, most notably
LHC, DAPHNE, PEP-II, {\it etc.}

An essential ingredient in our method is an expansion in gradients of a slowly
varying background. It is necessary to go beyond the leading order in
gradients, since the CP violating sources required by any baryogenesis 
mechanism influence particle dynamics only beyond the leading order.
The model whose dynamics we study here is very simple, a complex scalar field
with the lagrangian
\begin{equation}
{\cal L}= \partial_\mu\phi^\dag\partial^\mu\phi-\frac{1}{2}m^2\phi^\dag\phi
   + {\rm h.c.} + {\cal L}_{\rm int},
\label{c99.1}
\end{equation}
where for definiteness we shall take the quartic interaction
${\cal L}_{\rm int}=-\lambda(\phi^\dag\phi)^2\!/4$. The mass $m=m(x^\mu)$
represents coupling to a classical background field which varies in space and
time, an important example being the Higgs field condensate. Such a background
implies breakdown of the translational invariance, which has as a consequence
localization in position space, or equivalently, delocalization in momentum
space and breakdown of the quasiparticle picture in the plasma. This breakdown
is quite different from one related to the collective self-energy corrections.
Indeed, assuming a
planar symmetry, $m=m(z)$, which models the bubble wall at the electroweak
phase transition, we find that at the $p\,$th order in gradients the
momentum splits into $p+2$ branches ({\it cf.} figure 2), while the energy
remains conserved. A self-consistent description of the problem requires
$p+2$ distinct distribution functions,
or equivalently one distribution function and $p+1$ functions
that measure coherent quantum densities on phase space,
whose dynamics is described by generalized transport equations.
In the adiabatic limit 
(of frequent scattering) the coherent densities are suppressed, and one thus
recovers the quasiparticle picture with one semiclassical transport equation
for a distribution function which describes the dynamics on a
modified shell on phase space.

%
%
\section{The Kadanoff-Baym equations}
\label{sec: Kadanoff-Baym equations}

%
%
\begin{figure}
\leavevmode
\hspace{1truecm}
\epsfxsize=9truecm \epsfbox{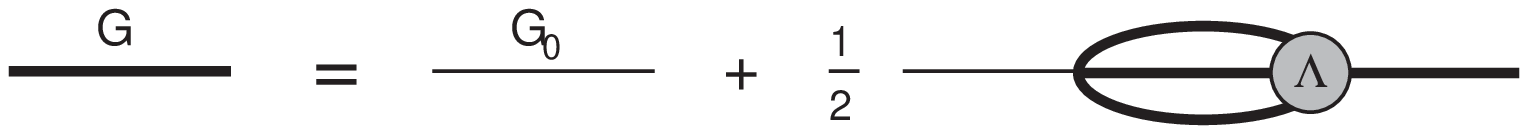}
\vspace{0.truecm}
\baselineskip 17pt
\caption{Diagrammatic representation of the Schwinger-Dyson equation
  for the scalar theory with ${\cal L}_{\rm int}=-\lambda(\phi^\dag\phi)^2/4$.
  $G$ and $G_0$ denote the full and free propagators, respectively,
  and $\Lambda$ the four point function.}
\end{figure}
  The Kadanoff-Baym equations (KBE) describe the out-of-equilibrium dynamics of
quantum fields, and can be derived from the Schwinger-Dyson equations (SDE)
in the Keldysh closed time contour (CTC) formalism
\cite{kb,dmrhh}.
In figure 1 we show the SDE for the scalar theory~(\ref{c99.1}).
In the weakly coupled limit the four point function can be approximated by
$\Lambda\propto \lambda$, so that the SDE for the propagator closes.
The self-energy $\Sigma$ is in this case computed in the Born approximation,
which is illustrated by the two loop diagram in figure 1.

  Since we are primarily interested in
the dynamics of the propagator in the presence of a slowly varying
background, we write the KBEs in the mixed (Wigner) representation
as follows \cite{jkp2}
\begin{eqnarray}
   &&
\cos{\Diamond}\DBR{\Omega^2\pm i\omega\Gamma}{G^{r,a}}
      = 1 \qquad \qquad \qquad {\rm (PE)}
\label{c99.2}
  \\
   &-& \sin{\Diamond }\left(\DBR{\Omega^2}{iG^{<,>}}
                     -\DBR{i\Sigma^{<,>}}{G_R}\right)
\nonumber\\
      &=& \frac{1}{2}\cos{\Diamond}\left(\DBR{\Sigma^>}{G^<}
           -  \DBR{\Sigma^<}{G^>} \right)\qquad {\rm (QBE)} ,
\label{c99.3}
\end{eqnarray}
where the operator
$\Diamond\{f\}\{g\} \equiv \frac{1}{2}\left[\partial_X f
\cdot \partial_k g - \partial_k f \cdot \partial_X g \right]$
is a generalized Poisson bracket.
The PE~(\ref{c99.2}) is the dynamical equation for the retarded and advanced
propagators, $G^{r,a}$, while the QBE~(\ref{c99.3}) is the transport equation
for the quantum Wigner functions, which in the coordinate representation read
\begin{eqnarray}
G^>(x,y)       &=& -i \langle \phi(x)\phi^\dagger(y) \rangle
\,,\qquad
G^<(x,y)       = -i \langle \phi^\dagger(y)\phi(x) \rangle .
\label{c99.4}
\end{eqnarray}
Various quantities in equations~(\ref{c99.2}) and~(\ref{c99.3}) are defined
as follows
\begin{eqnarray}
\Omega^2 &=& G_0^{-1}-\Sigma_R \equiv k^2-m^2-\Sigma_R
\nonumber\\
G^{r,a} &=& G_R\mp i{\cal A}
,\qquad
\Sigma^{r,a} =
  \Sigma_R\mp i\omega\Gamma ,
\label{c99.5}
\end{eqnarray}
where $\Sigma_R$ and $\Gamma$ denote the real part of the self-energy
and the damping rate, respectively,
${\cal A}$
denotes the spectral function defined by
\begin{equation}
{\cal A} \equiv \frac{i}{2}\left( G^r - G^a \right)
               = \frac{i}{2}\left( G^> - G^< \right),
\label{c99.6}
\end{equation}
while $G_R$ is the real part of the propagator, which is related to the
spectral function by the following spectral integral
$G_R=(1/\pi){\cal P} \int d\omega\prime {\cal A}(\omega^\prime)/
(\omega-\omega^\prime)$.
A similar spectral relation holds for $\Sigma_R$ and
$\omega\Gamma=i(\Sigma^>-\Sigma^<)/2$.
Finally, the weak coupling expansion ensures that
the self-energies $\Sigma^{<,>}$ may be computed in the Born approximation.

The mixed representation is defined by the Wigner transform,
which is the Fourier transform with respect to the relative
coordinate $r=x-y$. For example, for the propagator we have
\begin{equation}
   G(k;X) = \int d^4r e^{ik\cdot r} G(X+r/2,X-r/2),
\label{c99.7}
\end{equation}
where $X$ is the average coordinate defined by $X=(x+y)/2$.

Equations~(\ref{c99.2}) and~(\ref{c99.3}) represent an accurate description
of the plasma dynamics
background is slowly varying, {\it i.e.} when the gradient approximation
applies. Formally, this is the case when $\vert\vert\Diamond\vert\vert\ll 1$.
Physically, this condition holds when the de Broglie wave
length of excitations is small in comparison to
the characteristic length scale of the background variation.
For the electroweak phase boundary the characteristic scale
is the bubble wall thickness $L\sim (10-20)/T$ \cite{mp}, whereas for
plasma excitations $\partial_k \sim k^{-1}$, so that
$\vert\vert\Diamond\vert\vert\sim\vert\vert\partial_X\partial_k\vert\vert
\sim 1/kL$, implying that the gradient approximation is accurate
when $k \gg L^{-1}$, which is close to thermal equilibrium
satisfied by most of plasma excitations.

%
%

\section{Propagator equation}
\label{sec: Propagator equation}

We now consider the propagator equation~(\ref{c99.2}). For simplicity
we assume a planar symmetry such that $m=m(x)$, and set $\Sigma_R=0$.
For the electroweak phase transition this case corresponds to
a planar bubble wall in the wall frame. To second order in gradients
the propagator is solved by \cite{jkp1}
\begin{equation}
G^{r,a}\rightarrow G_{p=2}=\frac{1}{k_0^2-k_x^2}
   + \frac{1}{2}\frac{(m^2)^{\prime\prime}}{(k_0^2-k_x^2)^3}
   - \frac{1}{2}\frac{2k_0^2(m^2)^{\prime\prime}+[(m^2)^{\prime}]^2}
        {(k_0^2-k_x^2)^4} ,
\label{c99.8}
\end{equation}
where $k_0^2=\omega^2-k_{\small \parallel}^2-m^2(x)$ denotes the unshifted
pole, $(m^2)^{\prime\prime}\equiv \partial_x^2m^2$, and
$(m^2)^{\prime}\equiv \partial_x m^2$. Note that at second order in gradients
there is no pole shift, but instead multiple poles emerge. Now let us make
the following decomposition of the quantum Wigner functions
\begin{equation}
G^< = -2i{\cal A}n\,,\qquad G^> = -2i{\cal A}(n+1),
\label{c99.9}
\end{equation}
where $n$ is the generalized distribution function. Recall that the spectral
function ${\cal A}$~(\ref{c99.6}) defines the spectrum (poles) of the
propagator. For example, at leading order in gradients the spectral function
becomes a sharp on-shell projector:
${\cal A}\rightarrow {\cal A}_0
   = (\pi/2k_0)\left[\delta(k_x-k_0) + \delta(k_x+k_0)\right]$.
To understand how is this projective behaviour modified by the
presence of the derivative corrections in Eq.~(\ref{c99.8}),
we study the following spectral integral
\begin{equation}
{\cal I}_p[{\cal T}]
     = \frac{2}{\pi}\int_{-\infty}^\infty dk_x{\cal A}_p{\cal T}
\rightarrow \sum_{i=0}^{p+1}c_i{\cal T}^{(i)}(k_0) .
\label{c99.10}
\end{equation}
Note that the series terminates at the $p+1\,$st order in derivatives, where
$p$ is the highest order gradient in the propagator $G_{p}$.
For example, when $p=2$,
\begin{equation}
{\cal I}_2[n] = \frac{n(k_{\rm sc})}{k_{\rm sc}}
   + c_2 (\partial_{k_x}^2 n)(k_0)  + c_3 (\partial_{k_x}^3 n)(k_0) ,
\label{c99.11}
\end{equation}
where $k_x=k_{\rm sc}$ defines the semiclassical dispersion relation
\begin{equation}
k_{\rm sc} = k_0   + \frac{1}{8}\frac{(m^2)^{\prime\prime}}{k_0^3}
               + \frac{5}{32}\frac{[(m^2)^{\prime}]^2}{k_0^5}.
\label{c99.12}
\end{equation}
This agrees with the dispersion relation obtained by the standard WKB method
\cite{jkp1}. The additional higher order derivative terms in Eq.~(\ref{c99.11})
signify the breakdown of the
quasiparticle picture and emergence of quantum correlations, and they are
direct consequence of the multipole structure of $G_p$~(\ref{c99.8}).

%
%

\section{Transport equations}
\label{sec: Transport equations}

We now discuss the implications of the quasiparticle picture breakdown on
transport. To this end
we simplify the QBE~(\ref{c99.3})
by keeping only the higher order derivative term contributing to
the flow term \cite{jkp2} to obtain
\begin{equation}
   - (\Diamond - \Diamond^3/3) \DBR{\Omega^2}{iG^<}
      = \frac{1}{2}\left(\DBR{\Sigma^>}{G^<}
           -  \DBR{\Sigma^<}{G^>} \right) +o(\Diamond^2,\Gamma) .\quad
\label{c99.13}
\end{equation}
In order to make use of the decomposition (\ref{c99.9}) and the properties
of the spectral function ${\cal A}$, we shall perform spectral integrals
weighted by powers of the momentum over Eq.~(\ref{c99.13}). Upon defining
the moment functions
\begin{equation}
  f_l\equiv \frac{2}{\pi}\int_{-\infty}^\infty dk_x k_x^l{\cal A}_{p=2}n
,\qquad (l=0,1,2,..),
\label{c99.14}
\end{equation}
these integrals result in the following set of coupled transport equations
\begin{equation}
\omega\partial_t f_l +\tilde k_l\partial_xf_{l+1}
   -\frac{l}{2}(\partial_x k_0^2)f_{l-1}
   -\frac{l(l-1)(l-2)}{48}(\partial_x^3 k_0^2)f_{l-3}
  = {\rm Coll}_l  ,
\label{c99.15}
\end{equation}
where ${\rm Coll}_l$ is the collision term, which approximately has the
following form
\begin{equation}
  {\rm Coll}_l =-\Gamma^> f_l+\Gamma^<(f_l+k_0^{2-l}\kappa_l).
\label{c99.16}
\end{equation}
Here $\Gamma^{<,>}$ denote the damping rates computed in the Born
approximation \cite{jkp2} and, for example, $\kappa_0=k_{\rm sc}$.
To any finite order $p$ in the gradient expansion, there are only
$p+2$ independent moments and in particular at second order in gradients 
one can derive the constraint:
$f_4 - 4k_0f_3 + 6k_0^2f_2 - 4k_0^3f_1 + k_0^5f_0 =0$,
so that Eqs.~(\ref{c99.15}) close for the first four moments,
$l=0,1,2,3$.

The following picture has thus emerged.
Due to the higher order poles in the propagator~(\ref{c99.8}),
the quasiparticle dispersion relation has split into four branches
defined by
\begin{equation}
k_x\equiv \tilde k_l
     = k_0 + \frac{(l-1)(l-2)}{16}\frac{(m^2)^{\prime\prime}}{k_0^3}
  + \frac{l^2-5l+5}{32}\frac{[(m^2)^{\prime}]^2}{k_0^5},
\label{c99.17}
\end{equation}
such that for example $\tilde k_0=k_{\rm sc}$ in
Eq.~(\ref{c99.12}). This is illustrated in figure 2. Note that the
splitting diverges in the infrared, where gradient approximation breaks
down.  For a particle with an energy $\omega$ incoming onto the bubble wall,
the momentum splits into four branches, and the full description of the
dynamics then requires defining one distribution function on each branch.
%
%
\begin{figure}
\leavevmode
\hspace{3truecm}
\epsfxsize=5truecm \epsfbox{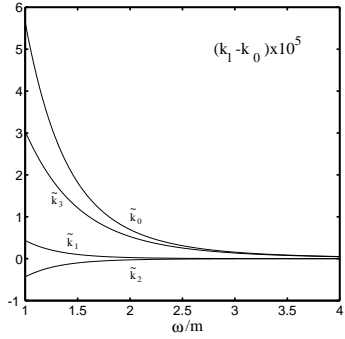}
\vspace{0.truecm}
\baselineskip 17pt
\caption{The hypersurfaces $k_x=\tilde k_l$
on which the moments $f_l$ flow according to Eqs.~(\ref{c99.15}-\ref{c99.16}).}
\end{figure}
%

%
%

\section{Quantum coherence and semiclassical limit}
\label{sec: Quantum coherence and semiclassical limit}

  We shall now briefly describe an alternative formulation of the
transport equations~(\ref{c99.15}-\ref{c99.16}), which is convenient
for discussion of the semiclassical limit. Let us first define
\begin{eqnarray}
  f &=& k_{\rm sc} f_0 , \qquad\qquad\quad\;\;\,
   f_{\rm qc1} = f_1- f
\nonumber\\
    f_{\rm qc2} &=& f_2/\kappa_2 - f_{\rm qc1} - f , \qquad
     f_{\rm qc3} = f_3/k_0^2 - f_{\rm qc2}- f_{\rm qc1} - f ,
\label{c99.18}
\end{eqnarray}
which then implies the following transport equation for $f$:
\begin{equation}
\omega \partial_tf + k_{\rm sc} \partial_x(f+f_{\rm qc1})
   =  -\Gamma^> f - \Gamma^<(f+1) .
\label{c99.19}
\end{equation}
The distribution function $f$ measures population density of
particles on the hypersurface $k_x=\tilde k_0\equiv k_{\rm sc}$, while the
densities $f_{\rm\! qci}$ measure the coherent quantum correlations
between neighboring hypersurfaces $k_x=\tilde k_{i-1}$ and $k_x=\tilde k_i$
defined in Eq.~(\ref{c99.17}). Eq.~(\ref{c99.19}) is already in the form of
the standard Boltzmann equation, apart from the presence of an unknown
function $f_{\rm\! qc1}$. Equations for $f_0$ and $f_1$ then imply
\begin{equation}
\omega(\partial_t +\Gamma_{\rm\! qc})f_{\rm\! qc1}
+\kappa_2-k_{\rm sc}\partial_x f_{\rm\! qc1}
+\partial_x\kappa_2(f_{\rm\! qc1}+f_{\rm\! qc2})
+\kappa_2\partial_xf_{\rm\! qc2}
=s_1,
\label{c99.20}
\end{equation}
where the source
\begin{equation}
s_1 = (k_{\rm sc}-\kappa_2)\partial_x f
+(\partial_x k_0^2/2k_{\rm sc})-\partial_x\kappa_2 f
\label{c99.21}
\end{equation}
represents coherent mixing of $f$ and $f_{\rm qc1}$. One can obtain
similar equations for the coherent quantum densities $f_{\rm\! qc2}$
and $f_{\rm\! qc3}$.
The coherent densities $f_{\rm\! qci}$ are all damped at the rate
$\Gamma_{\rm\! qc} \equiv (\Gamma^>-\Gamma^<)/\omega$, which is
an out-of equilibrium generalization of the on-shell damping rate
\cite{Weldon}.  Finally, in the frequent scattering limit defined as
\begin{equation}
\Gamma_{\rm\! qc}L\gg 1
\label{c99.22}
\end{equation}
one can neglect $f_{\rm cq1}$ in Eq.~(\ref{c99.19}) and one obtains
a semiclassical equation in which quasiparticles flow along modified
semiclassical trajectories \cite{jkp2}.

%
%

\section{Conclusions and outlook}
\label{sec: Conclusions and outlook}

  We have presented a self-consistent derivation of transport equations
for scalar theory beyond the leading order in gradients. We found that
as a consequence of localization in space, the quasiparticle
picture of transport breaks down. We have shown how to self-consistently
reformulate transport theory by including information about the varying
background to a finite order in gradients. Our result are the transport
equations~(\ref{c99.15}-\ref{c99.16}), which at second order
in gradients describe the dynamics on four momentum branches in phase space.
We have also briefly mentioned that one recovers the semiclassical picture
in the (adiabatic) limit of frequent scattering.

  The question which we are addressing now is how to generalize our results to
mixing scalar fields and massive chiral fermions, which are
the cases relevant for baryogenesis studies.

\section*{References}
%
%
\nc{\anap}[3]  {{\it Astron.\ Astrophys.\ }{{\bf #1} {(#2)} {#3}}}
\nc{\ap}[3]    {{\it Ann.\ Phys.\ }{{\bf #1} {(#2)} {#3}}}
\nc{\apj}[3]   {{\it Ap.\ J.\ }{{\bf #1} {(#2)} {#3}}}
\nc{\apjl}[3]  {{\it Ap.\ J.\ Lett.\ }{{\bf #1} {(#2)} {#3}}}
\nc{\app}[3]   {{\it Astropart.\ Phys.\ }{{\bf #1} {(#2)} {#3}}}
\nc{\araa}[3]  {{\it Ann.\ Rev.\ Astron.\ Astrophys.\ }{{\bf #1} {(#2)} {#3}}}
\nc{\arnps}[3] {{\it Ann.\ Rev.\ Nucl.\ and Part.\ Sci.\ }{{\bf #1} {(#2)}
{#3}}}
\nc{\ijmp}[3]  {{\it Int.\ J.\ Mod.\ Phys.\ }{{\bf #1} {(#2)} {#3}}}
\nc{\ijtp}[3]  {{\it Int.\ J.\ Theor.\ Phys.\ }{{\bf #1} {(#2)} {#3}}}
\nc{\jmp}[3]   {{\it J.\ Math.\ Phys.\ }{{\bf #1} {(#2)} {#3}}}
\nc{\mpl}[3]   {{\it Mod.\ Phys.\ Lett.\ }{{\bf #1} {(#2)} {#3}}}
\nc{\nat}[3]   {{\it Nature }{{\bf #1} {(#2)} {#3}}}
\nc{\ncim}[3]  {{\it Nuov.\ Cim.\ }{{\bf #1} {(#2)} {#3}}}
\nc{\np}[3]    {{\it Nucl.\ Phys.\ }{{\bf #1} {(#2)} {#3}}}
\nc{\pr}[3]    {{\it Phys.\ Rev.\ }{{\bf #1} {(#2)} {#3}}}
\nc{\prl}[3]   {{\it Phys.\ Rev.\ Lett.\ }{{\bf #1} {(#2)} {#3}}}
\nc{\pl}[3]    {{\it Phys.\ Lett.\ }{{\bf #1} {(#2)} {#3}}}
\nc{\prep}[3]  {{\it Phys.\ Rep.\ }{{\bf #1} {(#2)} {#3}}}
\nc{\phys}[3]  {{\it Physica\ }{{\bf #1} {(#2)} {#3}}}
\nc{\rmp}[3]   {{\it Rev.\ Mod.\ Phys.\ }{{\bf #1} {(#2)} {#3}}}
\nc{\rpp}[3]   {{\it Rep.\ Prog.\ Phys.\ }{{\bf #1} {(#2)} {#3}}}
\nc{\sjnp}[3]  {{\it Sov.\ J.\ Nucl.\  Phys.\  }{{\bf #1} {(#2)} {#3}}}
\nc{\spjetp}[3]{{\it Sov.\ Phys.\ JETP }{{\bf #1} {(#2)} {#3}}}
\nc{\yf}[3]    {{\it Yad.\ Fiz.\ }{{\bf #1} {(#2)} {#3}}}
\nc{\zetp}[3]  {{\it Zh.\ Eksp.\ Teor.\ Fiz.\ }{{\bf #1} {(#2)} {#3}}}
\nc{\zp}[3]    {{\it Z.\ Phys.\ }{{\bf #1} {(#2)} {#3}}}
\nc{\ibid}[3]  {{\sl ibid.\ }{{\bf #1} {(#2)} {#3}}}
%
%

\end{document}